\documentclass[aps,pre,preprint,superscriptaddress]{revtex4-1}
\usepackage{graphicx}
\usepackage{dcolumn}
\usepackage{bm}
\usepackage[centertags]{amsmath}
\usepackage{amsfonts}
\usepackage{amssymb}
\usepackage{amsthm}
\usepackage{newlfont}
\usepackage{color}
\usepackage{natbib}
\usepackage{subfigure}

\begin{document}

\title{Jarzynski matrix equality: calculating free energy difference
by non-equilibrium simulations with an arbitrary initial distribution}
\author{Biao Wan}
\affiliation{School of Physical Sciences, University of Chinese Academy of Sciences, Beijing 100049 China}

\author{Cheng Yang}
\affiliation{School of Physical Sciences, University of Chinese Academy of Sciences, Beijing 100049 China}

\author{Yanting Wang}
\affiliation{State Key Laboratory of Theoretical Physics, Institute of Theoretical Physics, Chinese Academy of Sciences, Beijing 100190 China}

\author{Xin Zhou}

\email[To whom correspondence should be addressed. Electronic Mail: ]{xzhou@ucas.ac.cn}
\affiliation{School of Physical Sciences, University of Chinese Academy of Sciences, Beijing 100049 China}

\date{\today}

\begin{abstract}
The Jarzynski equality (JE), which relates works of non-equilibrium trajectories to the free energy difference of the initial and final states of the non-equilibrium process, provides an efficient way to calculate free energies of systems in simulations and experiments. However, wider applications of the JE are limited by the requirement that the initial distribution of non-equilibrium trajectories must be equilibrium. Here we extend the JE to a matrix form, the Jarzynski matrix equality (JME), which transforms the free energies of metastable conformational regions in the initial system to that of final one. Therefore, we can calculate the free energies from non-equilibrium trajectories which started  from an arbitrary initial distribution. We demonstrate the application of the JME in toy models, Lennard-Jones fluids, and polymer chain models, show its good efficiency in calculation of free energy with a satisfactory accuracy. The JME  extends applications of the non-equilibrium method in estimate of free energy in complex system where the initial global equilibrium is difficult to reach.
\end{abstract}


\maketitle{}

\section{INTRODUCTION}
\label{sec:introduction}
In the past decades, an important progress in non-equilibrium statistical physics is the development of the fluctuation
theorems~\cite{Crooks1998, Crooks2000,Klages2013}, in particular, the Jarzynski equality (JE)~\cite{Jarzynski1997,jarzynski1997-pre} which relates the distribution of irreversible works of a non-equilibrium process to the difference of equilibrium free energy. The JE can be written as
\begin{equation}
\exp(-\beta\triangle A)=\langle \exp(-\beta W)\rangle,
 \label{JE}
\end{equation}
where $\beta=\frac{1}{k_{B}T}$ is the reciprocal of temperature multiplied by the Boltzmann constant $k_{B}$, $\triangle A$ is the free energy difference between the initial and final systems, and $\langle \cdots \rangle$ denotes the average over the ensemble of trajectories which starts from the equilibrium distribution of the initial system and ending with the final system. The work $W$ is defined as
\begin{equation}
W[x(t)]= \int_{0}^{\tau}\frac{\partial H(x(t),\Lambda(t))}{\partial\Lambda}\dot{\Lambda}dt,
\label{work}
\end{equation}
where $x(t)$ is a simple notation of a trajectory in conformational space 
within the time interval $[0,\tau]$. The trajectory can be followed a deterministic or a stochastic dynamic under the time-dependent Hamiltonian $H(x,\Lambda(t))$ with the special protocol $\Lambda(t)$. Here $\dot{\Lambda}$ denotes the time derivative of $\Lambda(t)$. Thus the work is a functional of non-equilibrium trajectory $x(t)$. In this paper, $x(t)$ represents the whole trajectory, and a particular conformation of the trajectory at time $t$ is denoted as $x_{t}$.

The JE provides a direct way to estimate the free energy difference in experiments, such as single-molecule pulling
~\cite{Hummer2001,Liphardt2002,Harris2007}, and simulations~\cite{Jarzynski2008,Hummer2005,Park2003,Jensen2002}. However, there is a major difficulty in applying JE. The right part of Eq.~(\ref{JE}), the ensemble average of work, is an exponential function dominated by rare trajectories with small work values. Therefore, inadequate sampling of these rare events could results in a biased estimation, as described by the Jensen inequality~\cite{jarzynski1997-pre}. One way to overcome this problem, in single-molecule pulling experiments, ones applied a very stiff spring potential to obtain a work distribution with an approximate
Gaussian~\cite{Park2004}. The exponential average of work then is estimated correctly even from inadequate samples~\cite{HummerJCP2001,Zuckerman2002,ZuckermanCPL2002,Gore2003}.
Some recent attempts have also been done by applying enhance sampling techniques to generate more small-work nonequilibrium trajectories~\cite{Jarzynski2001,Sun2003,Ytreberg2004,Chelli2013}, and improving simulation efficiency a little. 

Another essential difficulty in applying JE is to achieve an equilibrium conformational sample of the initial system as the starting points of non-equilibrium trajectories, which usually requires very (even impractical) long equilibrium simulation in the initial system, 
for example, in macroscopic biomolecules, it is usually very hard to reach equilibrium in normal simulation time scale, since the conformational space of the system consists of multiple long lifetime metastable states separated by high free energy barriers. 

Recently, some works were tried to extend the application of JE by not requiring the initial equilibrium distribution of trajectories. 
For example, Maragakis~\emph{et al.}~\cite{Maragakis2008} extended the Crooks theorem~\cite{Crooks2000} to calculate the free energy difference between two metastable conformational regions. Similarly, Junier~\emph{et al.}~\cite{Junier2009} derived a fluctuation relation under partial-equilibrium conditions to estimate the free energy branches of metastable states in single molecule experiments. Very recently, a theoretical extension of the JE for arbitrary initial distribution is also discussed by Gong and Quan~\cite{Quan2015}. However, these works usually apply the time-reverse process of non-equilibrium trajectories to reimburse the deviation of initial distribution from the equilibrium one, which limits their application in some cases.

In this paper, we extend the JE to be compatible with an arbitrary initial distribution and any non-equilibrium protocol  without requiring the strict time-reverse process. We present a practical form of the extended JE when the initial distribution is in a local equilibrium inside each of metastable conformational regions but not equilibrate among these regions. The local equilibrium is much easy to reach from any initial distribution after short-time local relaxation. 
An arbitrary initial distribution can easily evolute to the required local equilibrium inside each of metastable states by short relaxation, thus our method in practice can be applied for any arbitrary initial distribution~\cite{Huang2009,Gong2009}. The original JE now is replaced by a matrix equality, named as the Jarzynski matrix equality (JME), which connects free energies of metastable regions in the initial system to that in the final system of the non-equilibrium process.

The essential idea of the JME is formulized as below. The partition function of a system with multiple metastable states can be written as $Z=\sum_{\mu} Z_{\mu}$, where $Z_{\mu}$ denotes the local partition function of the $\mu$-state. Here, we suppose that all metastable states are included and boundaries of the states are neglected, \emph{i.e.}, the super-basin approximation~\cite{Miller1999}. When a non-equilibrium process changing the initial Hamiltonian of system to the final one, we have 
$Z_{\mu}(f) = \sum_{\nu} \pi_{\mu\nu}(f,i) Z_{\nu}(i)$, where $Z_{\mu}(f)$ and $Z_{\nu}(i)$ are the partition functions of metastable state $\mu$ in the final system $f$ and state $\nu$ in the initial system $i$, respectively. The transition matrix $\pi_{\mu\nu}(f,i)$ relates to works of trajectories which transition from the $\nu(i)$ to $\mu(f)$, \emph{i.e.}, 
$\pi_{\mu,\nu}(f,i) = T_{\mu\nu}(f,i) \langle \exp(-\beta W) \rangle_{\mu\nu}$.
Here $T_{\mu\nu}(f,i)$ is the transition probability of trajectories started from $\nu(i)$ to $\mu(f)$, and $\langle\cdot\rangle_{\mu\nu}$ is the ensemble average over all these transition trajectories. 

Particularly, if the final system is chosen to be identical to the initial one, the above relation is reduced to the linear equation $Z_{\mu}=\sum_{\nu}\pi_{\mu\nu}Z_{\nu}$, which can be used to solve $Z_{\mu}$ and thus the free energy of the whole system.  

\section{THEORY AND METHODS}
\label{sec:theory}

\subsection{Basic theory}
For any initial distribution $\rho_{init}(x_0)$, we can reproduce the equilibrium distribution of the initial 
Hamiltonian $H(x;\lambda_0)$, $\rho_{eq,0}(x_0) = \omega(x_{0}) \rho_{init}(x_0)$~\cite{Gong2009}, so we have,
\begin{equation}
\langle\omega(x_{0}) \delta(x - x_{\tau}) \exp(-\beta W[x(t)])\rangle = \frac{\exp(-\beta H(x; f))}{Z(i)},
 \label{newJE}
\end{equation}
where $\langle \cdots \rangle$ is the ensemble average over trajectories which started from the initial distribution $\rho_{init}(x_{0})$, and $H(x; f)$ is the final Hamiltonian. Eq.~(\ref{newJE}) is a direct extension of the JE with an arbitrary initial distribution. It is also an extension of the formula given by G. Hummer and A. Szabo~\cite{Hummer2001}. The original JE can be obtained by integrating both sides of Eq.~(\ref{newJE}) with respect to $x$~\cite{Jarzynski2008} after putting $\omega(x_{0})=1$. Due to too wide histogram of the values of $\omega(x_0)$, the equation is usually not helpful to directly apply in calculation of free energy, except in very low dimension (such as one-dimension, or two dimension) cases. 

\subsection{Local equilibrium approximation}
Usually, the conformational space of a system can be divided into many metastable regions (states),  
and the local equilibrium inside each of them can reaches in a short time scale, while the equilibrating between states interstate needs longer time. 
Therefore, within a short simulation time, any initial distribution can relaxes to a locally-equilibrating one which is in proportion to the equilibrium distribution inside each of metastable conformational regions (states). In other word, 
the weighting function $\omega(x_0)$ in Eq.~(\ref{newJE}) is approximately a constant $\omega_{\nu}$ for all $x_0$ inside the state ${\nu}$, but the value of $\omega_{\nu}$ varies with respect to the index of state $\nu$~\cite{Gong2009}. 
Thus, we almost always partition the whole conformational space into some metastable states. 
The summation of these states approximately equals to the whole conformational space, since these states do not overlap each other and the contribution of boundaries of these states are ignorable. 

This approximation greatly simplifies Eq.~\ref{newJE} in practical application. 
We define local partition function of metastable state, $Z_{\nu} = \int \Theta_{\nu}(x) e^{-\beta H(x)} dx$. Here 
$\Theta_{\nu}(x_0)$ is the characteristic function of the state $\nu$, which is unity if $x$ is inside the state, otherwise. Thus, $Z_{\nu} \propto p_{\nu}$, where $p_{\nu}$ is the equilibrium probability of visiting the state $\nu$. 
Thus the partition function $Z$ of the system approximately equals to the sum of all local partition functions $Z =\sum_{\nu} Z_{\nu}$, or write in a matrix form, $\mathbf{Z}=(Z_{1},\ldots,Z_{\mu},\ldots,Z_{n})^{T}$. We have
\begin{equation}
  \sum_{\nu} \langle \omega_{\nu} \Theta_{\nu}^{i}(x_0) \delta(x - x_{\tau}) \exp(-\beta W[x(t)]) \rangle = \frac{\exp(-\beta H(x; f))}{Z(i)},
  \label{local}
\end{equation}
where the indices $i$ and $f$ represent the initial and final Hamiltonians, respectively. By multiplying the state-characteristic function of the final system, $\Theta_{\mu}^{f}(x)$, on both sides of Eq.~(\ref{local}) and integrating the equation for $x$, we have
\begin{equation}
  \int dx \sum_{\nu} \Theta_{\mu}^{f}(x) \omega_{\nu} \langle \Theta_{\alpha}^{i}(x_0) \delta(x - x_{\tau}) \exp(-\beta W[x(t)])\rangle = \frac{Z(f)}{Z(i)}\int dx\Theta_{\mu}^{f}(x) \rho_{eq,f}(x).
  \label{transition}
\end{equation}
Here $\rho_{eq,f}(x)$ is the equilibrium probability function of the final system, which is equal to $\frac{1}{Z(f)} \exp[-\beta H(x; f)]$. Eventually, we have the main expression of the JME, 
\begin{equation}
 \sum_{\nu}\pi_{\mu\nu}(f, i) Z_{\nu}(i)=Z_{\mu}(f),
 \label{newrelation}
\end{equation}
where $\pi_{\mu\nu}(f,i) = \frac{n_{\mu\nu}}{n_{\nu}(0)} \frac{1}{n_{\mu\nu}} \sum_k \exp(-\beta W_k)=T_{\mu\nu}(f,i)\frac{1}{n_{\mu\nu}} \sum_k \exp(-\beta W_k)$. Here $n_{\nu}(0)$ is the total number of trajectories started from state $\nu$ in the initial system, $n_{\mu\nu}$ is the number of trajectories started from $\nu$ and ended in state $\mu$ in the final system. The summation of $k$ is limited in the transition trajectories from $\nu$ to $\mu$, and $W_k$ is the work along the transition trajectory $k$. $T_{\mu\nu}(f,i)= \frac{n_{\mu\nu}}{n_{\nu}(0)}$ is the transition probability of non-equilibrium trajectories starting from state $\nu(i)$ and ending in state $\mu(f)$. For equilibrium processes, $W_k =0$, Eq.~\ref{newrelation} becomes the normal detailed balance condition.

\subsection{Loop protocol and linear equation}
If we choose a loop nonequilibrium protocol for which the initial Hamiltonian is identical to the final one,
\emph{i.e.}, $\Lambda(t_{0})=\Lambda(\tau)$, Eq.~\ref{newrelation} becomes a linear equation, $\sum_{\nu}\pi_{\mu\nu}Z_{\nu}=Z_{\mu}$, or
 \begin{equation}
 \mathbf{\Pi}\mathbf{Z}=\mathbf{Z},
 \label{matrix}
\end{equation}
which offers a practical way to estimate the local partition functions $\{Z_{\mu}\}$. Since the matrix elements are nonnegative, $\mathbf{\Pi}$ is a positive and non-reductive matrix. According to the Perron-Frobenius theorem, a positive and non-reductive matrix  has only one eigenvector whose all components are the same sign, corresponding to the eigenvalue with maximal module of the matrix. Therefore $\mathbf{\Pi}$ has unique non-negative vector $\mathbf{Z}$, relating to  the eigenvalue  $\lambda=1$ in Eq.~(\ref{matrix}). 
In principle, the corresponding eigenvalue should be unity. In practice, however, due to numeric and statistical errors, it might slightly differ from unity. The deviation of the eigenvalue from unity is a good criterion to determine if the generated non-equilibrium trajectories are sufficient.

\section{RESULTS AND ANALYSIS}
\label{sec:results}
We  applied the JME to three kinds of systems: one-dimensional multiple-well potentials, Lennard-Jones (LJ) fluids near the liquid-solid coexistence condition, and a polymer chain with closing and opening end-end states.
The implementation of the JME usually contains four steps: (i) choose conformations in each metastable state and relax them shortly under the initial Hamiltonian to serve as initial conformations at $t=0$; (ii) start the non-equilibrium simulations with the loop protocol $\Lambda(0) = \Lambda(\tau)$ from the prepared initial conformations; (iii) calculate the matrix elements of $\mathbf{\Pi}$, and its unique same-sign eigenvector and the corresponding eigenvalue $\lambda$. The eigenvector gives $\mathbf{Z}$ with an arbitrary constant coefficient. Here if $\lambda$ is approximately equal to $1$ provides a criterion of the calculation.

\subsection{1D toy models}
\begin{figure}[h]
  \centering
  \subfigure[]{
    \label{fig:1:a}
    \includegraphics[width=0.4\textwidth,angle=0]{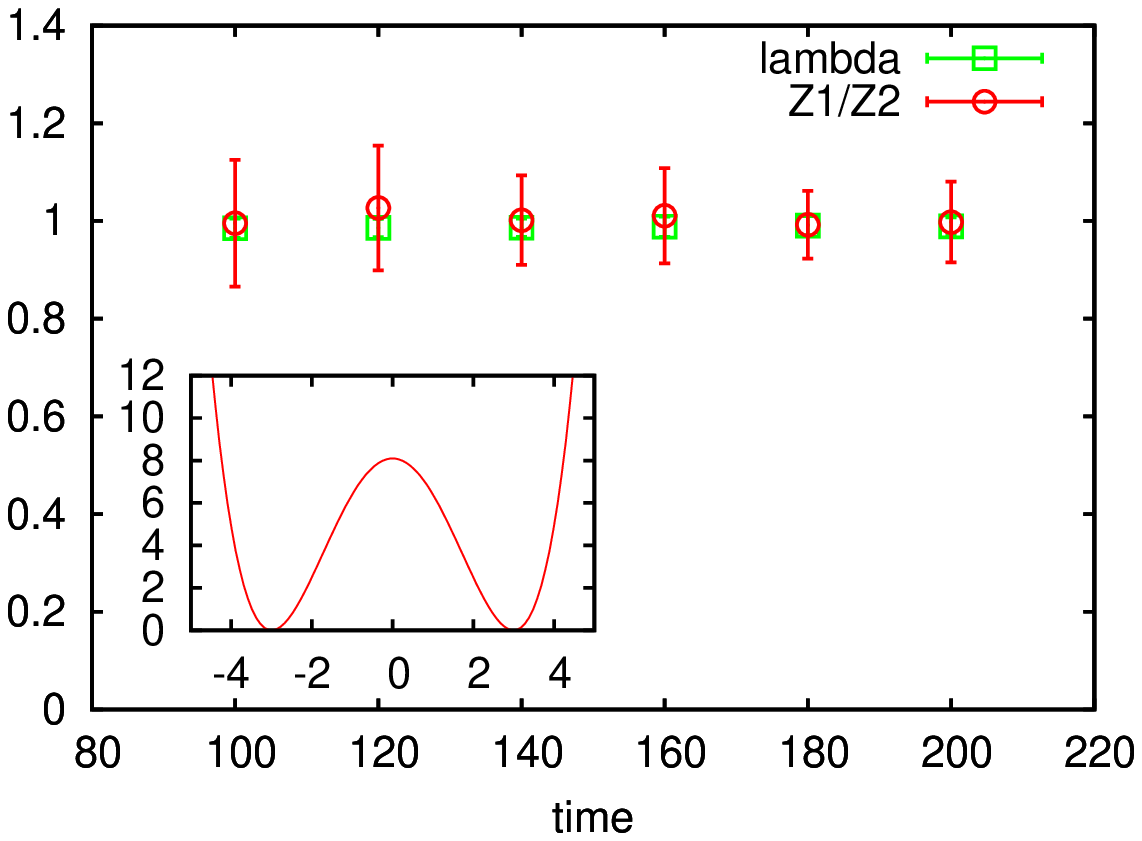}}
  \subfigure[]{
  \label{fig:1:b}
  \includegraphics[width=0.4\textwidth,angle=0]{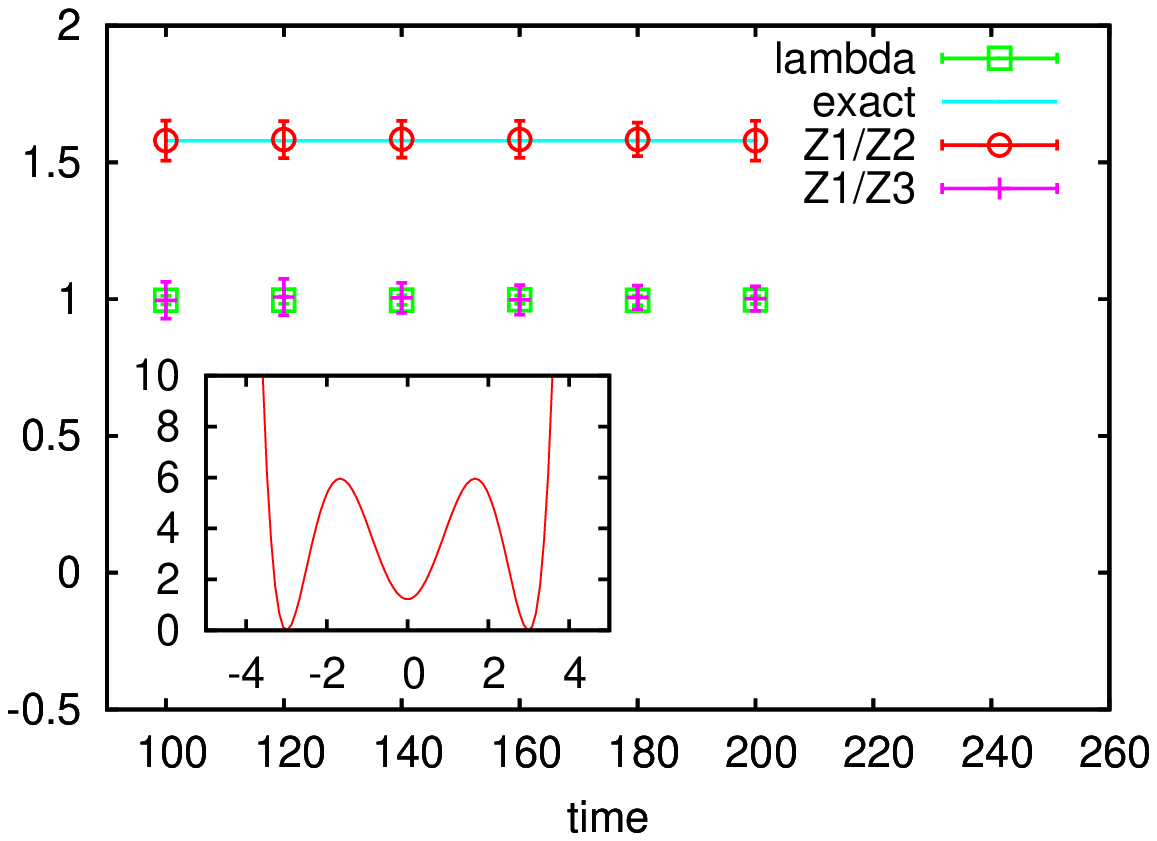}}
  \caption{Non-equilibrium simulation results of the one-dimensional symmetric multiple-wells. (a)The simulation results for the double-well potential. The $X$-axis denotes the simulation duration of a single trajectory.
The green symbols denote the eigenvalues of the transition matrix, and the red ones denote the partition function ratios
 of state 1 over state 2. The inset depicts the potential landscape. (b)The simulation results for the triple-well potential. The green symbols denote the eigenvalues of the matrix,
the red ones denote the partition function ratios of state 1 over state 2, and the burgundy ones denote the ratio
of state 1 over state 3. The inset depicts the corresponding potential.}

  \label{fig:1}
\end{figure}

We first consider a simple one-dimensional symmetric multiple-well potential (plotted in Fig.~\ref{fig:1:a})
\begin{equation}
 U=\frac{1}{2}k(q^{2}-9)^{2}.
 \label{twowell}
\end{equation}
A particle moving under this potential is simulated according to the over-damped Langevin dynamics with $\beta=1$. The mobility of the particle $1/\gamma$  is 0.2. A simple non-equilibrium loop protocol linearly changing the parameter $k$ with time from $k(0)=0.2$ to $k(\tau)=0.2$, is applied, namely, $k(t)=0.2-0.36 t/\tau$ when $0<t\leq \frac{\tau}{2}$, and $k(t)=0.02+0.36 t/\tau$ when $\frac{\tau}{2}<t\leq \tau$.

We ran six sets of non-equilibrium simulations with different simulation durations of $\tau=100,120,140,160,180$, and $200$, respectively. For each set, we totally generated $2000$ trajectories, and the initial allocation of these trajectories in the two states was that the left potential well had $1200$ trajectories and the right well had $800$ trajectories. Each trajectory initially went through a short equilibrium simulation to reach local equilibrium distribution inside its initial potential well. The results are shown in Fig.~\ref{fig:1:a}. The green symbols denote the eigenvalues of the transition matrix $\mathbf{\Pi}$, and the red ones denote the partition function ratios of state 1 over state 2, whose ideal values should both be $1$. As expected, the values of the two parameters calculated from different simulation durations are all close to $1$.
The error bars in the figure are the standard deviations obtained by repeating the simulations and calculations $100$ times.

We also applied the method to a one-dimensional triple-well potential
\begin{equation}
 U=\frac{1}{2}k(q^{2}-9)^{2}(q^{2}+0.3),
 \label{tripletwell}
\end{equation}
where we set $k=0.1$ and $\beta=1$. Mobility of the particle is $0.2$. The three states are labeled as 1, 2, and 3 from left to right in Fig.~\ref{fig:1:b}. The exact partition 
function values are $Z_{1}=1.58 Z_{2}$ and $Z_{1} = Z_{3}$. The nonequilibrium protocol is similar to the previous example. We ran six sets of nonequilibrium simulations with different durations of $\tau=100,120,140,160,180$,
and $200$, respectively. For each set, we generated $3000$ trajectories evenly
distributed in the three states to estimate the matrix elements. The results are shown in Fig.~\ref{fig:1:b}. The green symbols denote the eigenvalues of the transition matrix, whose ideal value should be $1$; the red ones denote the partition function ratios of state 1 over state 2, whose ideal value should be $1.58$; and the burgundy ones denote the ratios of state 1 over state 3, whose ideal value should be $1$. The values calculated from the trajectories with different simulation lengths are all close to their ideal values. The error bars in the figure are the standard deviations obtained by repeating the simulations and calculation $100$ times.

\subsection{Lennard-Jones liquid-solid coexistence}
In this subsection, we apply the JME to a 32-particle Lennard-Jones (LJ) system inside a cubic box with the side length $11.8\AA$ with the periodic boundary condition. Noose-Hover thermostat. The pairwise LJ interaction 
\begin{equation}
U=4\epsilon(\frac{\sigma^{12}}{r^{12}}-\frac{\sigma^{6}}{r^{6}}).
\label{LJ}
\end{equation}
Here the cut-off radius is $5.8 \AA$ smaller than the half of simulation box length. The nonequilibrium loop protocol uniformly changes the parameter $\epsilon$ in $40$ steps from $0.34$ $\mathrm{kcal/mol}$ to $0.35$ $\mathrm{kcal/mol}$, then from $0.35$ $\mathrm{kcal/mol}$ to $0.33$ $\mathrm{kcal/mol}$, and finally from $0.33$ $\mathrm{kcal/mol}$ to $0.34$ $\mathrm{kcal/mol}$. The system temperature was fixed at $T = 55$ $\mathrm{K}$, employing the Noose-Hover thermostat, where the liquid and solid phases coexist. The non-equilibrium work for each trajectory is  $W=\sum_{i}(\triangle_{i}\epsilon)4(\frac{\sigma^{12}}{r_{i}^{12}}-\frac{\sigma^{6}}{r_{i}^{6}})$, where $i$ goes over the $40$ steps, and $r_{i}$ denotes the distance between two ends at the terminal  of  $i^{\mathrm{th}}$ step. The fraction of the two coexistence phases can be calculated from non-equilibrium simulations by applying Eq.~\ref{newrelation}. To distinguish the liquid and solid phases, we employed the local bond order parameters measuring the local structure around a particle~\cite{Auer2004}, defined as

\begin{equation}
  q_{lm}(i)=\frac{1}{N_{b}(i)}\sum_{j=1}^{N_{b}(i)}Y_{lm}(\tilde{r}_{ij}),
  \label{LBOP}
\end{equation}

where $\tilde{r}_{ij}$ is the unit vector from particle $i$ to particle $j$, the summation goes over all neighboring particles
$N_{b}(i)$ of particle $i$, and $Y_{lm}(\tilde{r}_{ij})$ is the spherical harmonic function, with $l$ and $m$ taking integer values of $l=0,1,\cdots,$ and $m = -l, \cdots, l$. Specifically, $q_{6}$ is known as a good order parameter for distinguishing the liquid and solid phases~\cite{Auer2004}. In simple liquids, there are no preferred orientations around a particle and thus the structural correlation decays rapidly. In contrast, for particles in a solid-like environment the vectors are correlated:
\begin{equation}
 \mathbf{q_{6}(i)}\cdot\mathbf{q_{6}(j)}=\sum_{m=-6}^{6}q_{6m}(i)\cdot q^{\ast}_{6m}(j).
 \label{Q6}
\end{equation}
where the asterisk indicates the complex conjugate. The average of the correlation functions provides a rough criterion
for distinguishing the liquid phase and the solid phase as~\cite{Russo2012}
\begin{equation}
s=\frac{1}{32}\sum_{i}\sum_{j=1}^{N_{b}(i)}\frac{\mathbf{q}_{6}(i)\cdot\mathbf{q}_{6}(j)}{|q_{6}(i)||q_{6}(j)|},
\label{order}
\end{equation}
In our MD simulations of this system, the value of the parameter changes continuously from $0$ to $12$ when it
evolves from the liquid phase to the solid phase. We regard the system as in the liquid phase when $s < 7$ and solid otherwise.

\begin{figure}[h]
  \centering
  \includegraphics[width=0.6\textwidth]{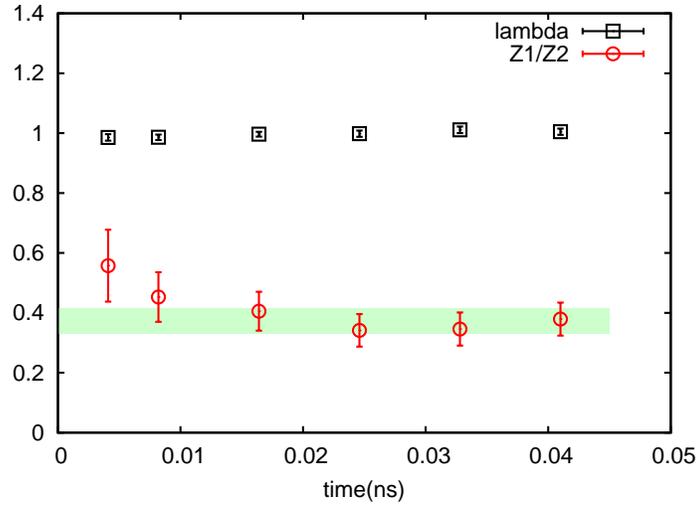}
  \caption{Nonequilibrium simulation results of the LJ system. The $X$-axis is the simulation duration of a single trajectory.
The black symbols  denote the eigenvalues of the matrix, and the red ones denote the partition function ratio of the
solid state over the liquid state. As a reference, the mean value from 20 sets of 400 equilibrium simulations (green shadow) is 0.37.}
  \label{fig:2}
\end{figure}

Figure.~\ref{fig:2} shows six sets of non-equilibrium simulations with the simulation length of each trajectory ranging from
$\tau=4.8$ to $42$ \textrm{ps}. For each set, we generated $400$ trajectories with half starting from liquid and the other half from solid.  The boot-strap method~\cite{Efron1993} was used to estimate errors. Each trajectory first went through a $0.2$  $\textrm{ps}$ simulation for initially local equilibration. The positive eigenvalues of the transition matrixes are found to approximately equal to $1$ due to the small work fluctuations in our simulations. The ratio of the solid-state partition function to the liquid-state one obtained from $15$-ps trajectories is already in good agreement with the reference value. The reference values and its error bars in Fig.\ref{fig:2} were obtained from $20$ $\times$ $400$  trajectories of $1$ nanosecond  each, shown as the green shadow in Fig.~\ref{fig:2}.

\subsection{Opening/closing polymer chain}
\begin{figure}[h]
  \centering
  \subfigure[]{
    \label{fig:3:a}
    \includegraphics[width=0.3\textwidth,angle=0]{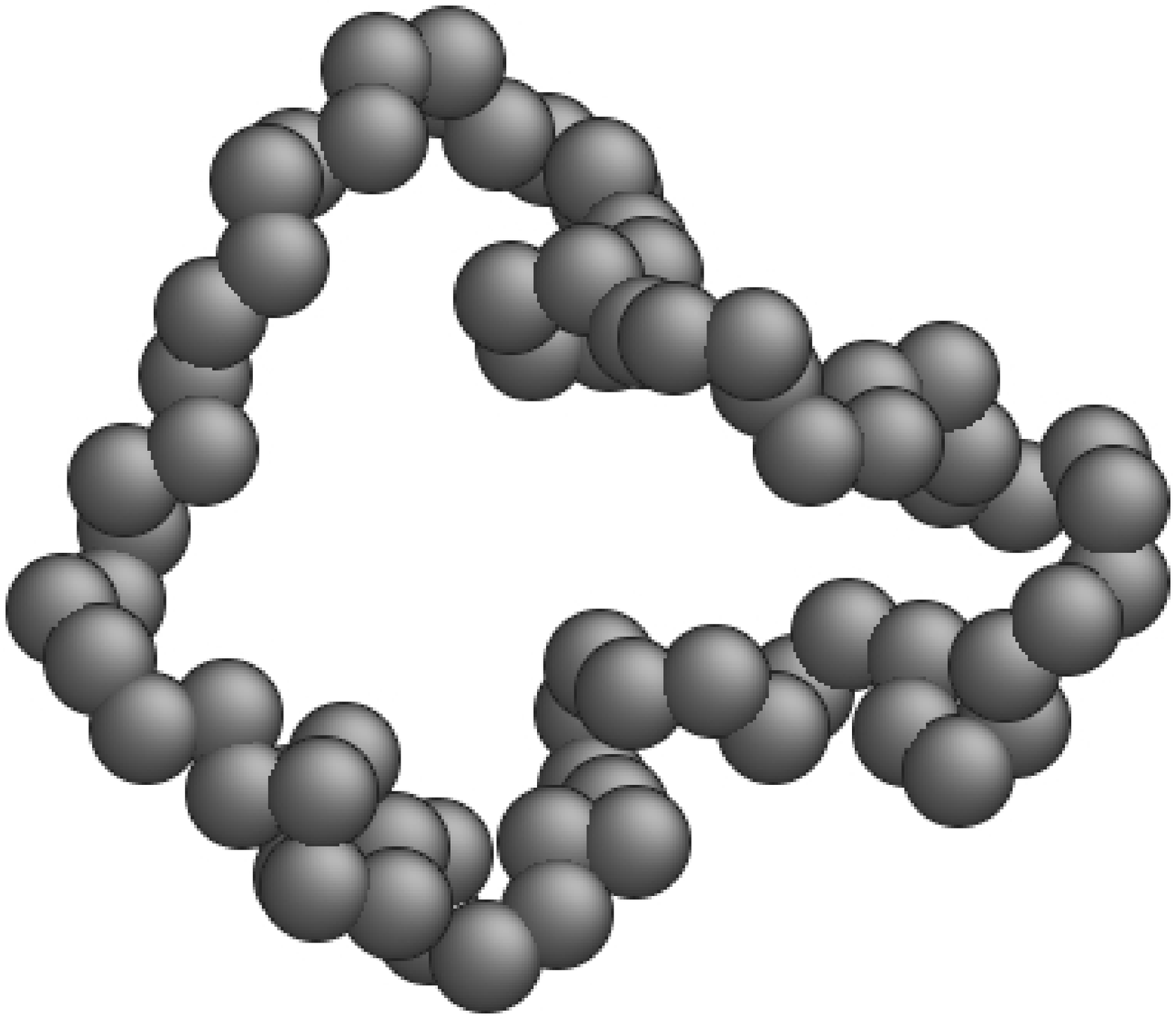}}
  \subfigure[]{
  \label{fig:3:b}
  \includegraphics[width=0.3\textwidth,angle=0]{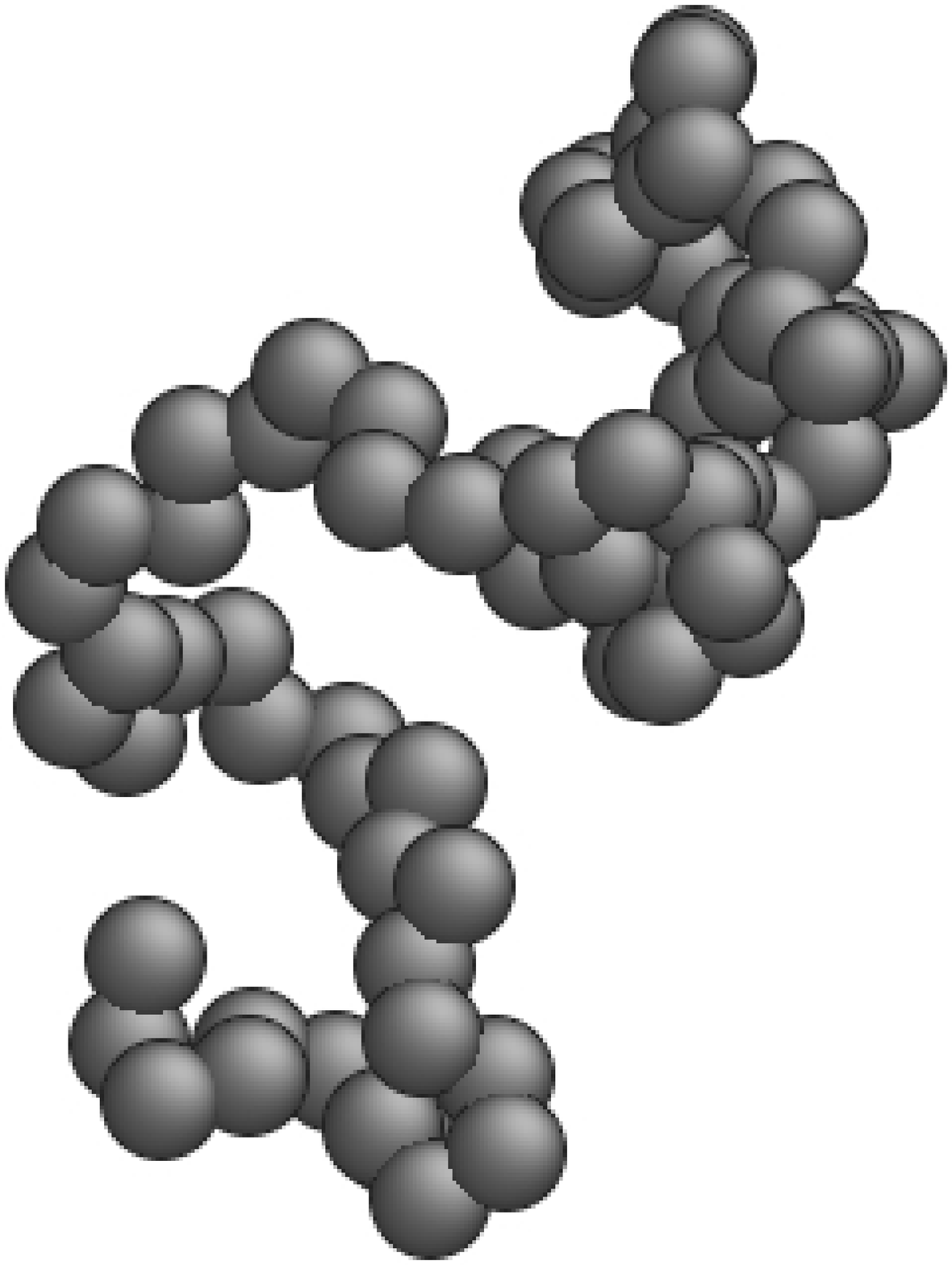}}
\subfigure[]{
  \label{fig:3:c}
  \includegraphics[width=0.6\textwidth,angle=0]{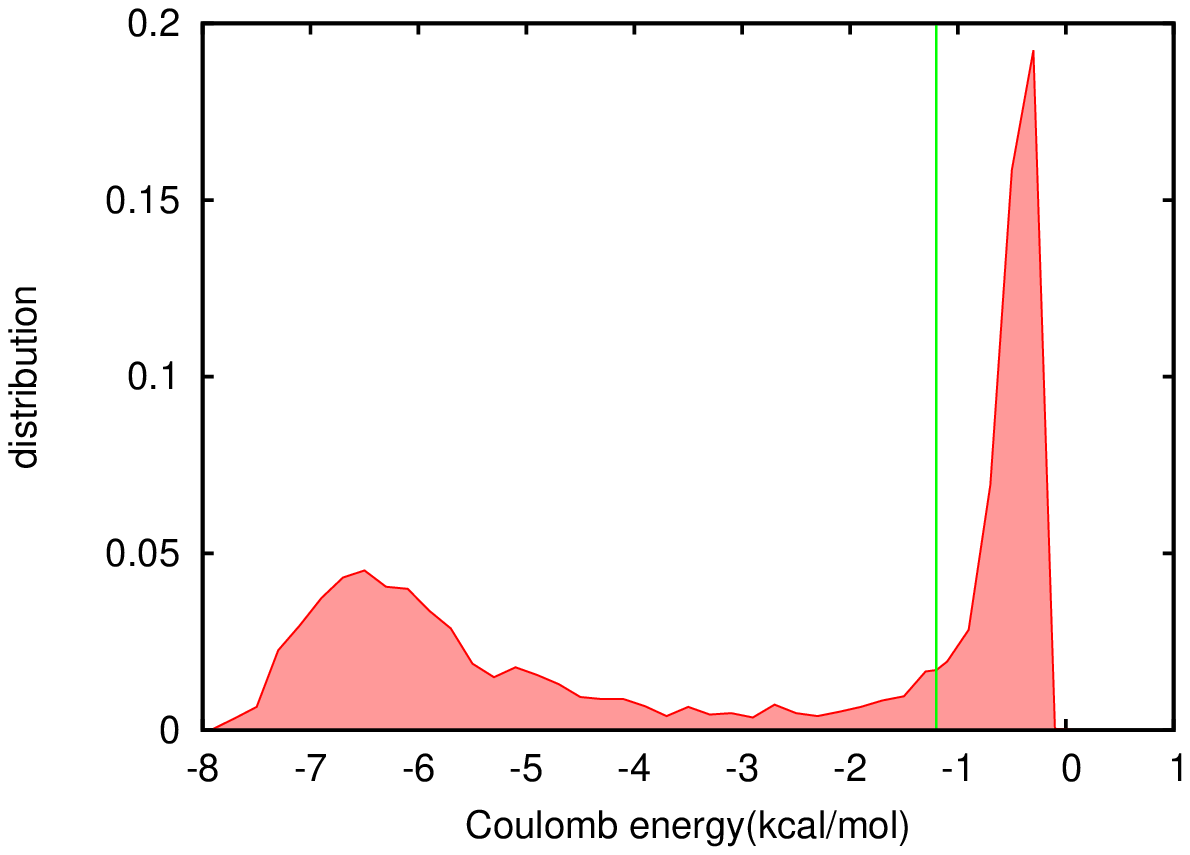}}
 \caption{(a) and (b) show an end-end closing  configuration and an end-end opening configuration of the polymer chain system, respectively.
(c) Coulomb energy distribution from 5000 samples shows  two well separated peaks: the energy greater than $-1.2$ belongs to the end-end opening state and the energy less than $-1.2$ belongs to the end-end closing configuration.}
  \label{fig:3}
\end{figure}

\begin{figure}[h]
  \centering
  \includegraphics[width=0.6\textwidth]{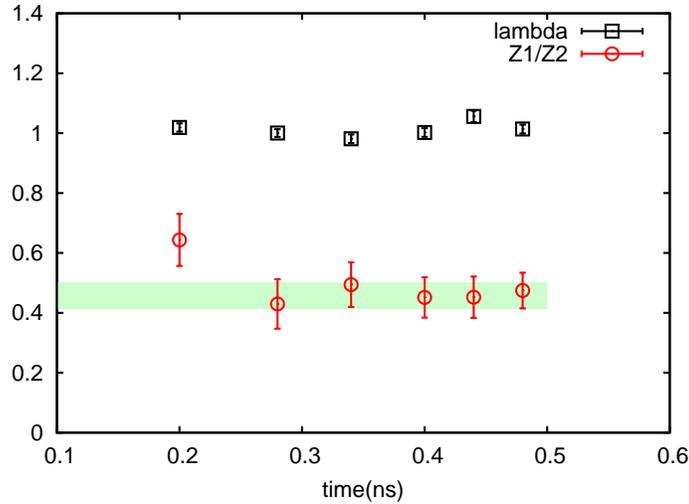}
  \caption{Non-equilibrium simulation results for the polymer chain model. The black symbols denote the eigenvalues of the matrix, the red ones denote the partition function ratio of the open state over the loop state. As a reference,
the mean value of 20 sets of 500 $2.5$-\textrm{ns} equilibrium simulations (green shadow) is 0.45.}
  \label{fig:4}
\end{figure}

 Next we study a model polymer chain consisting of $70$ atoms without periodic boundary condition. The atoms interact each other by the pairwise LJ potential and the neighbor atoms along the chain are connected by a harmonic spring  potential. In addition, we applied the Coulomb interaction to the two ending atoms of the chain to adjust the ratio of two metastable states, namely, the end-end closing and opening states, shown in Fig.~\ref{fig:3:a} and Fig.~\ref{fig:3:b}, respectively. The distribution of Coulomb energy shown in Fig.~\ref{fig:3:c} has two well separated peaks, which allows it to serve as an order parameter to distinguish the open and close states. The Coulomb potential is
\begin{equation}
  U_{e}(r)=-\frac{Q_{1}Q_{2}}{4 \pi \epsilon r} =\frac{\alpha}{r} ,
  \label{Coulomb}
\end{equation}
where $Q_{1}$ and $Q_{2}$ are the charges of the two ending atoms, $\epsilon$ is the dielectric constant. The cut-off radius of Coulomb interaction is $50\AA$. Since the Coulomb interaction controls the  opening-closing transition, the simplest way of enhancing the transition is to manipulate the coefficient $\alpha$. 
The non-equilibrium protocol consists of $20$ steps, during which the parameter $\alpha$ 
was changed from $1/34$ to $1/40$ and then back to $1/34$. The non-equilibrium work for each trajectory is $W=\sum_{i}-\frac{\triangle_{i} \alpha}{r_{i}}$, where $i$ goes over the $20$ steps, and $r_{i}$ denotes the distance between two ends at the $i^{\mathrm{th}}$ step. Fig.~\ref{fig:4} shows six sets of non-equilibrium simulations with the length of each trajectory ranging from $\tau=0.2$ to $0.48 \mathrm{ns}$. The system temperature was fixed at $T = 300$ $\mathrm{K}$, employing the Langevin thermostat. For each set of simulations, $500$ trajectories were generated and $250$ trajectories each state initially. At the beginning, each trajectory went through a $2$-ps simulation to reach the local equilibrium inside each state. The reference values and its error bars in  Fig.~\ref{fig:4} were obtained from $20 \times 500$ trajectories with $2.5 \mathrm{ns}$ each.
As shown in Fig.~\ref{fig:4}, very short non-equilibrium simulations ($0.28 \mathrm{ns}$ for each trajectory) are already sufficient to provide satisfactory results.

\begin{figure}[h]
  \centering
  \includegraphics[width=0.6\textwidth]{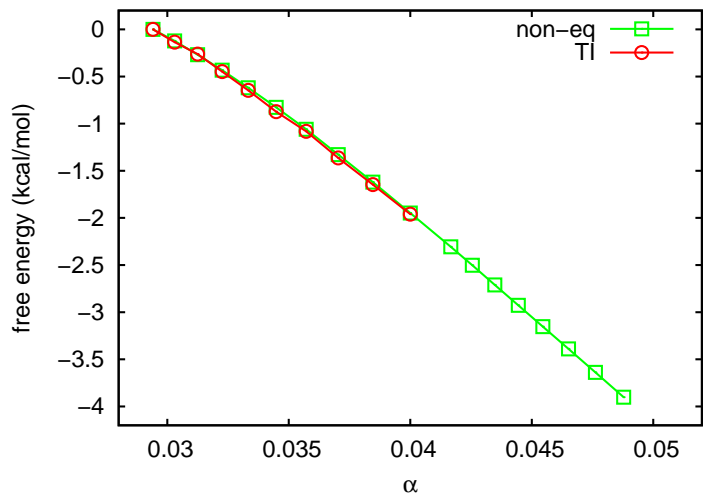}
  \caption{Free energy difference as a function of the Coulomb interaction strength $\alpha$ changing from 1/34 to 1/20. The green lines are the free energy as a function of $\alpha$  from the nonequilibrium simulation
by using Eq.(\ref{free-energy}) with an initial ratio of 0.43. As a reference, the red ones are the free energy as a function of $\alpha$ in [1/34,1/25] calculated by the thermodynamic integration method.}
  \label{fig:5}
\end{figure}

According to Eq.~(\ref{newrelation}), the free energy difference between initial and final systems can be calculated by
\begin{equation}
  \Delta A \equiv A(f) - A(i) = - k_{B} T \ln \frac{ \sum_{\mu\nu} \pi_{\mu\nu}(f,i) Z_{\nu}(i) } { \sum_{\nu} Z_{\nu}(i) },
  \label{free-energy}
\end{equation}
which is nothing else but the JE. A direct estimation of $\Delta A$ by the JE requires the initial system to be in the global equilibrium, in order to obtain an approximated equilibrium estimator,  probably $500$ trajectories of at least $1 \mathrm{ns}$ each are needed, though here we used a series of $500$ trajectories of $2.5 \mathrm{ns}$ each to get the reference value and its error bar.  In the JME, however, only local equilibrium initial conformations inside each state are required, which can be obtained in $1 \mathrm{ns}$ totally and $2 \mathrm{ps}$ each trajectory. Hence, the sum of simulation time of all non-equilibrium trajectories used to estimate global equilibrium is around $150 \mathrm{ns}$.
Once the local partition functions of two states in the system are obtained, Eq.~(\ref{free-energy}) can be applied.
For this system, $250$ non-equilibrium trajectories with the protocol changing $\alpha$ from $1/34$ to $1/20$ with $17$ intermediate values were applied, so the total simulation time of all trajectories is $42.5 \mathrm{ns}$. In Fig.~\ref{fig:5}, the green lines denote the calculated free energy as a function of $\alpha$. For comparison, we also calculated the free energy with $\alpha$ in $[1/34, 1/25]$ by the thermodynamic integration (TI) method based on equilibrium simulations~\cite{frenkel2001understanding}. We can see that the free energy profile obtained from the non-equilibrium simulation is in good agreement with the value obtained by the TI method.

\section{CONCLUSIONS AND DISCUSSION}
\label{discussion}
In comparison with the JE, the major benefit of the JME is that it does not require a global equilibrium initial distribution, thus can be applied in complex systems where the initial equilibrium distribution is hard to achieve.  
In addition, in the JME, trajectories are classified into some groups according to their beginning and ending metastable states, and the exponential average of works is estimated separately in each group. 
Since the fluctuation of works inside each group should not be larger than that of all trajectories, the grouping average might bring a little benefit in getting good estimate of the exponential average of works. 

\begin{figure}[h]
  \centering
  \includegraphics[width=0.6\textwidth]{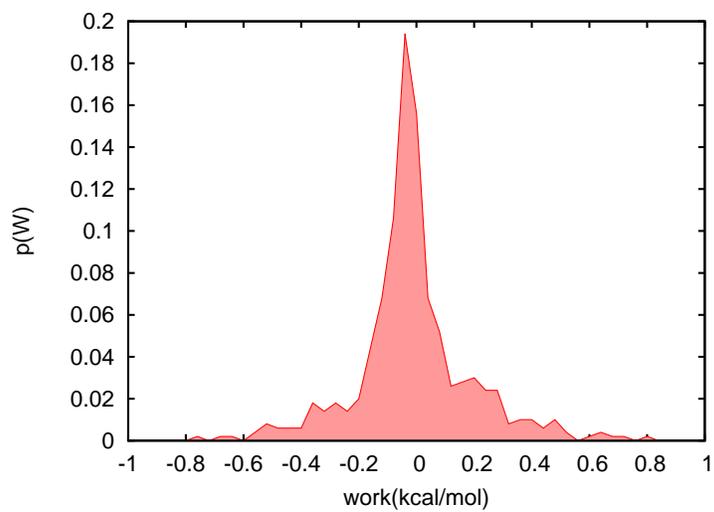}
  \caption{Work distribution of the fastest nonequilibrium simulation for the polymer chain model.
  }
  \label{fig:6}
\end{figure}

In the JME, it is a key to properly choose the non-equilibrium process to effectively enhance the transitions among metastable states, and to limit the work fluctuations in each group of non-equilibrium trajectories within the order of $k_{B} T$ at the same time so that the exponential averages of works can be good estimated. 
For the two purposes, the applied time-dependent biased potential in the JME should be designed to enable to decrease the free energy barriers separated important metastable states a few $k_{B} T$. In previous works, a similar bias potential was applied during the whole equilibrium simulation to enhance transitions then to estimate free energies based on the Bennett-Chandler approach~\cite{frenkel2001understanding,Montero1997,Chandler1978}. Whereas, this kind of bias potential is slowly applied and removed in the non-equilibrium JME simulations to enhance the transitions but without bring not too large work fluctuation. 
Therefore, it is a key to design the range-limited biased potential mainly on transition regions between metastable states, which usually require some \emph{a priori} understanding on metastable states, order parameter or reaction coordinates.   For example, in the polymer chain model, all our chosen non-equilibrium protocols produce small work fluctuations and apparently enhance transitions. The work fluctuation even in the fastest non-equilibrium simulation ($\tau = 0.2 {\mathrm ns}$), shown in Fig.~\ref{fig:6}, is still in the order of $1$ $\mathrm{k_{B}T}$, which guarantees the accuracy of free energy calculations.


For the systems studied in this paper, we assume all the metastable states are known and easy to distinguish. Therefore we can focus on the transition-related regions and design suitable biased potential to adjust the potential energy surface in the regions. In more general cases, when the metastable states of the system are unknown, we may combine the JME with our previously developed technique, the re-weighting ensemble dynamics method~\cite{Gong2009}, to obtain a more general formula for the JME without explicitly identifying metastable states and the transitions among them. This work is in progress~\cite{Yang2015}. 

\begin{acknowledgments}
Funding by the National Natural Science Foundation of China (Nos. 11175250, 91227115 and 11121403) and the Open Project from State Key Laboratory of Theoretical Physics are gratefully acknowledged. X.Z. also thanks the financial support from the Hundred Talent Program of the Chinese Academy of Sciences and fruitful discussion with D. P. Landau.
\end{acknowledgments}

\

\newpage

\end{document}